%% file: whel_preprint.tex
\documentclass[preprint,showpacs,preprintnumbers,amssymb]{revtex4}

\usepackage[english]{babel}
\usepackage{graphicx}
\usepackage{dcolumn}
\usepackage{epsfig}
\usepackage{setspace} 
\usepackage{bm}

\begin{document}


\title{\boldmath Helicity of the $W$ Boson in Lepton+Jets $t\overline{t}$ Events}

\input{list_of_authors_r1.tex}
\date{\today}

\begin{abstract}
We examine properties of $t\overline t$ candidates events in lepton+jets final states to establish the helicities of the $W$ bosons in $t\rightarrow W+b$ decays. Our analysis is based on a direct calculation of a probability that each event corresponds to a $t\bar t$ final state, as a function of the helicity of the $W$ boson. We use the 125 events/pb sample of data collected by the D\O\ experiment during Run I of the Fermilab Tevatron collider at $\sqrt{s}$=1.8 TeV, and obtain a longitudinal helicity fraction of $F_0$=0.56$\pm$0.31, which is consistent with the prediction of $F_0$=0.70 from the standard model. 
\end{abstract}

\pacs{14.65.Ha, 12.15.Ji, 12.60.Cn, 13.88.+e \hspace{2cm} FERMILAB-Pub-04/057-E} 
\maketitle

The observation of the top quark at the Fermilab Tevatron collider \cite{d0,cdf} has provided a new opportunity for examining detailed implications of the standard model (SM). In fact, the large mass of the top quark has led to speculation that its interactions might be especially sensitive to the mechanism of electroweak symmetry breaking and new physics that is expected to appear at the TeV energy scale. Several pioneering studies of the decays of the top quark have already appeared in the literature \cite{cdfwhel,spincorr}. Although these have been limited by small size of the data sample of the 1992-1996 Run I of the Tevatron collider, they have indicated nevertheless that it is feasible to measure subtle properties of the top quark predicted by the SM.

In this letter we report a measurement of the longitudinal component of the helicity of $W$ bosons from $t\rightarrow Wb$ decays in $t\bar{t}$ candidate events. The helicity of the $W$ boson is reflected in the
    angular distribution of the products of its decay. The analysis is based on a method of extracting parameters that was particularly effective for the measurement of the mass of the top quark \cite{massPRLnew,theses}.

An important consequence of a heavy top quark is that, to good approximation, it decays as a free quark. Its expected lifetime is approximately 0.5$\times10^{-24}$ $s$, and it therefore decays about an order of magnitude faster than the time needed to form bound states with other quarks \cite{khoze}. Consequently, the spin information carried by top quarks is expected to be passed directly on to their decay products, so that production and decay of top quarks provides a probe of the underlying dynamics, with minimal impact from gluon radiation and binding effects of QCD \cite{khoze,germans}.

The standard top quark decays through a V--A charged-current weak interaction. The emitted $b$ quark can be considered as essentially massless compared to the top quark ($m_b << m_t$). To conserve angular momentum, the spin of the $b$ quark, with its dominantly negative helicity (i.e., spin pointing opposite to its line of flight in the rest frame of the top quark) can therefore point either along or opposite to the spin of the top quark. In the first case, the spin projection of the vector $W$ boson must vanish (i.e., the $W$ is longitudinally polarized, or has zero helicity $W_{0}$). If the spin of the $b$ quark points opposite to the top quark spin, the $W$ boson must then be left-hand polarized (have negative helicity $W_{-}$). Hence, for massless $b$ quarks, a top quark can decay only to a left-handed or a longitudinal $W$ boson. In the SM, assuming $m_b=0$, the decay to longitudinal $W$ bosons is determined by the mass of the top quark and of the $W$ boson, and has a branching ratio \cite{kane}:

\begin{equation}
F_0=B(t\rightarrow W_{0}b) = \frac{m_{t}^{2}}{m_{t}^{2}+2M_{W}^{2}} = 0.70 \pm 0.01
\end{equation}
where the mass of the top quark is taken as $m_t$ = 174.3 $\pm$5.1 GeV/$c^2$ and of the $W$ boson as $M_W$ = 80.4 GeV/$c^2$ \cite{PDG}. (The impact of the finite $m_b\approx$ 4 GeV/$c^2$ on $F_0$ is negligible.)

To examine the nature of the $tbW$ vertex, we use $t \bar t$ candidates observed at the  D\O\ experiment \cite{d0det} in $p \bar p$ collisions at a center-of-mass energy $\sqrt{s}$=1.8 TeV. The data correspond to an integrated luminosity of 125 events/pb, and this analysis is based on the same lepton+jets sample that was used to extract the mass of the top quark in a previous D\O\ publication \cite{massPRD5}. That is, the signal is based on one of the $W$ bosons decaying into $l$+$\nu_{l}$, with $l$=$e$ or $\mu$, and the other $W$ decaying to two quarks ($q\bar q^\prime$); this leads to a final state characterized by one lepton and at least four jets (two from the fragmentation of the $b$ quarks). Making use of information contained in these events and comparing each individual event with the differential cross section for $t \bar t$ production and decay, we extract the fraction $F_0$ of longitudinal $W$-boson production in the data, assuming no contribution from right-handed $W$ bosons. In particular, we rely on a direct comparison of data to the matrix element for the production and decay of $t \bar t$ states \cite{massPRLnew,theses}. This method offers the possibility of increased statistical precision by using the decay of both $W$ bosons in these events, and is similar to that suggested for $t \bar t$ dilepton decay channels, and used in previous mass analyses of dilepton events \cite{dgk10}. A similar approach was also suggested for the measurement of the mass of the $W$ boson at the LEP collider at CERN \cite{berends}.  

An initial set of selection criteria was used to improve the acceptance for lepton+jets from $t\bar t$ events relative to background \cite{massPRD5}. These requirements were: $E^{\rm lepton}_T>20$ GeV, $|\eta_{e}|<2$, $|\eta_{\mu}|<1.7$, $E_T^{\rm jets}>15$ GeV, $|\eta_{\rm jets}|<2$, $\not\!\! E_T$$>20$ GeV, $|E_T^{\rm lepton}|+ $$\not\!\! E_T$$>60$ GeV, and $|\eta_{{\rm lepton+}\not\!\! E_T}|<2$. (Where $\eta$ and $E_T$ denote pseudorapidities and transverse energy of the lepton or jets, and $\not\!\! E_T$ the imbalance in transverse energy in the event.) A total of 91 events remained after imposing these requirements \cite{massPRD5}. The present analysis uses events that contain only four reconstructed jets (see below). 

The probability density for $t \bar t$ production and decay in the $e$+jets final state, for given value of $F_0$, is defined as:

\begin{eqnarray}
\label{ttbar}
P_{t \bar t}(F_0)&=&\frac{1}{12 \sigma_{t \bar t}} \int {\rm d}\rho_1 {\rm d}m_1^2 {\rm d}M_1^2
 {\rm d}m_2^2 {\rm d}M_2^2 \\ \nonumber
              &\times& \sum_{\mbox{perm},\nu}
 |{\cal M}_{t \bar t}(F_0)|^2 \frac{f(q_1)f(q_2)}{|q_1||q_2|} \Phi_6
W_{\rm jets}(E_{\rm p},E_{\rm j})
\end{eqnarray}
where ${\cal M}_{t \bar t}$ is the leading-order (LO) matrix element, $f(q_1)$ and $f(q_2)$ are the CTEQ4M parton distribution functions for the incident quarks \cite{cteq}, $\Phi_6$ is the phase-space factor for the 6-object final state, $\sigma_{t\bar t}$ is the total cross section for the LO $t\bar t$ production process, and the sum is over all twelve permutations of jets (the effective permutation of the indistinguishable jets from the decay of the $W$ was performed through a symmetrization of the matrix element) and all possible longitudinal momenta for neutrino solutions in $W$ decay. The integration variables used in the calculation are the two top quark invariant masses ($m_{1,2}$), the $W$ boson invariant masses ($M_{1,2}$), and the energy of one of the quarks from $W$ decay ($\rho_1$). Observed electron momenta are assumed to correspond to those of produced electrons. The angles of the jets are also assumed to reflect the angles of the partons in the final state, and we ignore any transverse momentum for the incident partons. These assumptions, together with energy and momentum conservation, introduce 15 $\delta$-functions in the integration of the probability density, and reduce the dimensionality of the remaining integrations to the five given in Eq. \ref{ttbar}. $W_{\rm jets}(E_{\rm p},E_{\rm j})$ corresponds to a function that parameterizes the mapping between parton-level energies $E_{\rm p}$ and jet energies measured in the detector $E_{\rm j}$. About 100,000 Monte Carlo (MC) $t \bar t$ events (generated with masses between 140 and 200 GeV/$c^2$ using {\sc herwig} \cite{herwig}, and processed through the D\O\ detector-simulation package) were used to determine $W_{\rm jets}(E_{\rm p},E_{\rm j})$. For $\mu$+jets final state, $W_{\rm jets}$ is expanded to include the known muon momentum resolution and an integration over muon momentum is added to Eq. \ref{ttbar}.

All processes that contribute to the observed final state must be included in the probability density. The final probability density is therefore written as:
\begin{equation}
P(x;F_0)={\rm c}_1P_{t\bar t}(x;F_0)+{\rm c}_2P_{bgd}(x)
\label{sumprob}
\end{equation}
where $c_1$ and $c_2$ are the signal and background fractions, and $x$ is the set of variables needed to specify the measured event. $P_{t\overline t}$ and $P_{bgd}$ refer to the signal and background production and decay probabilities, respectively. $W$+jets production contributes about 80\% to the background. The remainder of the background arises from multijet production where one jet mimics an electron. The {\sc vecbos} \cite{vecbos} $W$+jets matrix element is used to calculate the background probability density, which is integrated over the energy of the four partons that lead to jets, and over the $W$-boson mass, and summed over the 24 jet permutations and neutrino solutions. With the selections that we have used, the character of the multijet background is quite similar to that of $W$+jets, and we have therefore used {\sc vecbos} to also represent this component of the background, and have estimated a systematic uncertainty resulting from this assumption \cite{theses}.(Similarly, we have ignored the $\approx$ 10\% contribution to $t\overline t$ production from $gg$ fusion, and used only the $q\overline q\rightarrow t\overline t$ in ${\cal M}_{t \bar t}$.)

Effects such as geometric acceptance, trigger efficiencies, event selection, etc., are taken into account through a multiplicative function $A(x)$ that is independent of $F_0$. This function relates the $t \bar t$ and $W$+jets probability densities to their respective measured probability densities $P_m(x;F_0)$, as follows:
\begin{equation}
P_{m}(x;F_0)=A(x)[{\rm c}_1P_{t\bar t}(x;F_0)+{\rm c}_2P_{bgd}(x)]
\label{acc}
\end{equation}
Because the method involves a comparison of data with a leading-order matrix element for the production and decay process, we have restricted the analysis to events with exactly four jets, reducing the data sample from 91 to 71 events. To increase the purity of signal, a selection is applied on the probability of an event corresponding to background ($P_{bgd}$). This selection was used in Ref. \cite{massPRLnew,theses} to minimize a bias introduced by the presence of background, and it yields a sample of only 22 events. The selected cutoff value of probability density is based on MC studies carried out before applying the method to data, and, for a top quark mass of 175 GeV/$c^2$, it retains 71$\%$ of the signal and 30$\%$ of the background \cite{massPRLnew,theses}.

The probabilities are inserted into a likelihood function for $N$ observed events. The $t \bar t$ probability density contains contributions from both $W_{0}$ ($F_0$) and $W_{-}$ ($F_-$) helicities, and the ratio of $F_0/F_-$ is allowed to vary. The best estimate of $F_0$ is obtained by maximizing the following likelihood function with respect to $F_0$, subject to the constraint that $F_0$ must be physical, i.e., 0$\le F_0\le$1, and $F_-+F_0$=1 \cite{theses}:  
\begin{equation}
L(F_0)= e^{-N\int P_m(x,F_0) {\rm d}x} \prod_{i=1}^N P_m(x_i,F_0)
\label{like}
\end{equation}
where $P_m$ is the probability density for observing that event.

Inserting Eq. \ref{acc} into Eq. \ref{like}, the likelihood, becomes:
\begin{eqnarray}
\label{tot_lik}
-{\rm ln} L(F_0) &=& - \sum_{i=1}^N \ln[c_1 P_{t \bar t}(x_i;F_0) + 
c_2 P_{bgd}(x_i)] \\ \nonumber
         &+& N {\rm c}_1 \int A(x) P_{t \bar t}(x;F_0) {\rm d}x +
            N {\rm c}_2 \int A(x) P_{bgd}(x) {\rm d}x
\end{eqnarray}
The above integrals are calculated using MC methods. In this case the acceptance $A(x)$ takes the values 1.0 or 0.0, depending on whether the event is accepted or rejected.  The best values of $F_0$ and the parameters $c_i$ are obtained from minimizing --ln$L(F_0)$ with respect to all three parameters.

\begin{figure}
\begin{center}
\epsfig{file=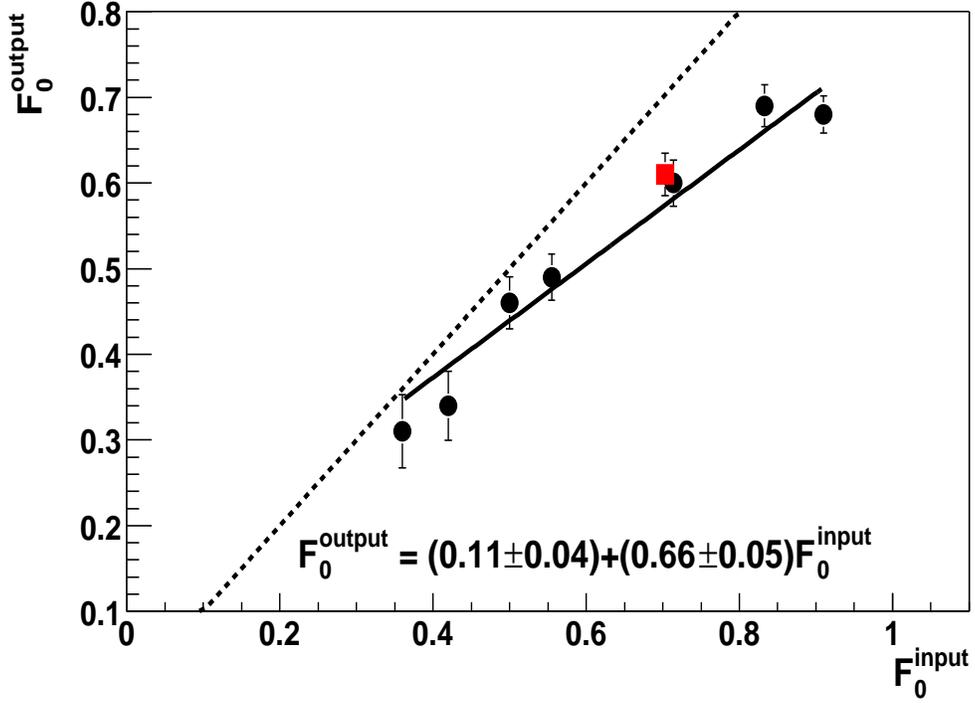,height=10cm,width=14cm}
\end{center}
\vskip -0.8cm
\caption[]{Result of $F_0$ extraction ($F_{0}^{output}$) as a function of $F_{0}^{input}$, for ensembles of 12 $t\bar t$ signal events and 10 $W$+jets for the {\sc pythia} samples (black dots) and the {\sc herwig} sample (square), after all selections. The dotted line has unit slope and passes through (0,0). The solid line is a fit to the results from {\sc pythia}.}
\label{line}
\end{figure}

\begin{figure}
\begin{center}
\epsfig{file=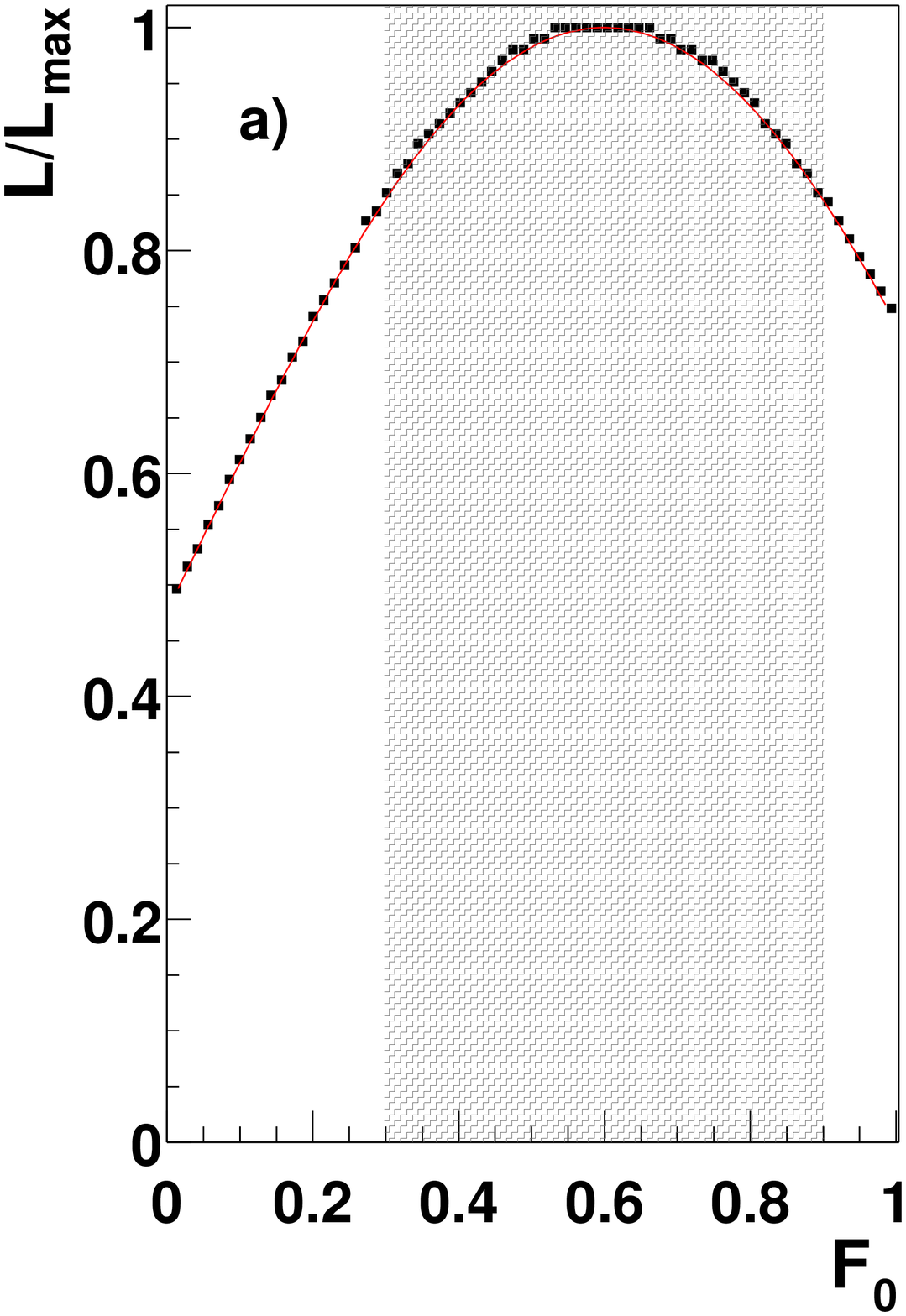,height=8cm,width=7cm}
\epsfig{file=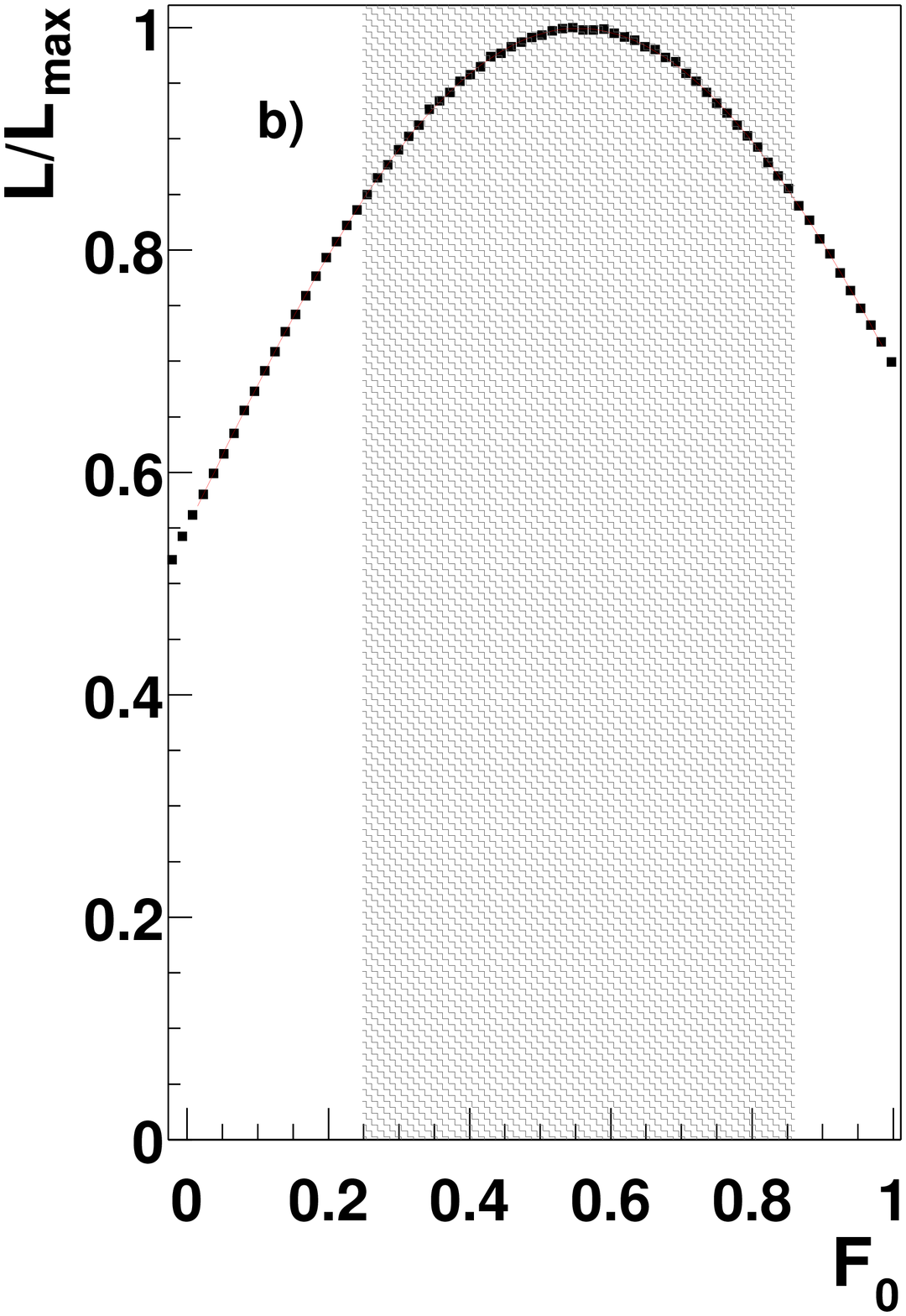,height=8.1cm,width=7.2cm}
\end{center}
\vskip -0.6cm
\caption[]{a) Likelihood normalized to its maximum value, as a function of $F_0$ for data from Run I. b) Likelihood as a function of $F_0$, after integration over $m_t$ (see text). The curves are 5th-order polynomials fitted to the likelihood.  The hatched area corresponds to the most narrow 68.27\% probability interval.}
\label{results}
\end{figure}

The response of the analysis to different input values of $F_0$ is examined by fluctuating the number of events according to a binomial distribution with an average of 12 events for signal ($S$) and 10 events for background ($B$). ($S/B=12/10$ was obtained in \cite{massPRLnew}.) Results from analyzing samples of {\sc pythia} MC \cite{pythia} events (shown in Fig. \ref{line}) indicate that a response correction must be applied to the data.  
Studies using resolution-smeared partons (rather than jets) indicate that the reason the response correction differs from unity  may have origin in gluon radiation, which is not included in our definition of probabilities.
We apply the correction from Fig. \ref{line} to the data, and Fig. \ref{results}a shows the result for the final sample of 22 events. For $m_t$=175 GeV/$c^2$, we find $F_0$=0.60$\pm$0.30(stat), and obtain a signal background ratio that is compatible with the value of 0.54 found in the mass analysis \cite{massPRLnew}.

\begin{figure}
\begin{center}
\includegraphics[width=.8\textwidth]{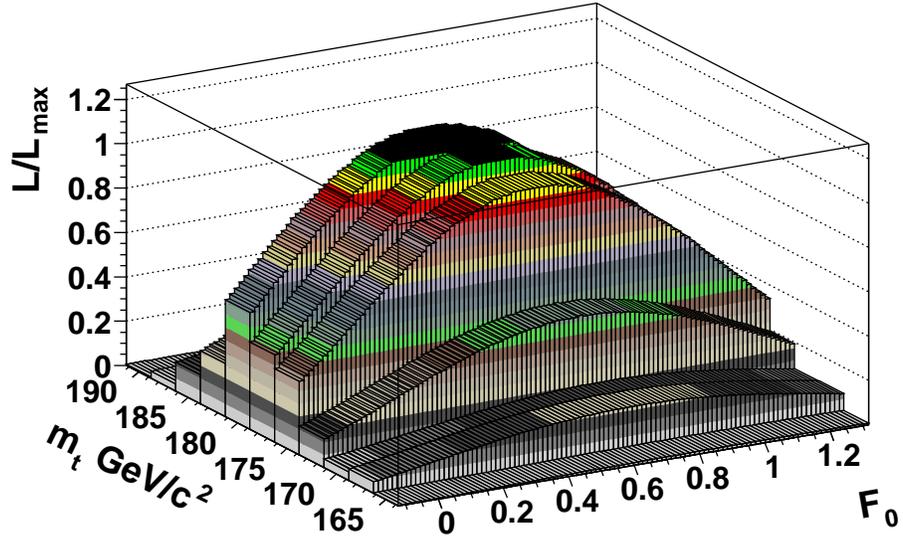}
\end{center}
\vskip -0.2cm
\caption[]{ Likelihood normalized to its maximum value as a function of $m_t$ and $F_0$.}
\label{topmass}
\end{figure}

\begin{figure}
\begin{center}
\includegraphics[width=.6\textwidth]{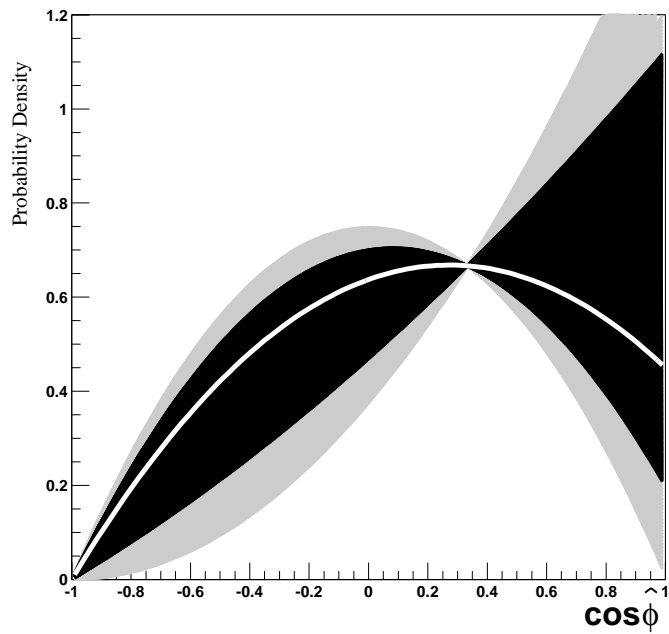}
\end{center}
\vskip -0.8cm
\caption{All possible decay functions cos $\hat \phi$ for different mixtures of $W_-$ and $W_0$ (grey region), where $\hat\phi$ refers to the decay angle in the $W$ rest frame. The result of our analysis, indicated by the black region, corresponds to the most probable value of $F_0$ and its 68.27$\%$ interval. The white line is the prediction of the SM.}
\label{result}
\end{figure}

When a probability density represents the data accurately, no systematic bias is expected in the extraction of any parameter through the maximum likelihood method. The current uncertainty in the top-quark mass is large enough to affect the value of $F_0$. For sufficiently high statistics, the likelihood can be maximized as a function of the two variables ($F_0$,$m_t$), which can then correctly take account of any correlations between the two parameters and the fact that $F_0$ is bounded between 0 and 1.  Given our limited statistics, the next best way to account for the uncertainty in $m_t$ is by projecting the two-dimensional likelihood onto the $F_0$ axis. In this way, the systematic uncertainty in $F_0$ from the uncertainty in $m_t$ can be obtained by integrating the probability over the mass, which we do from 165 to 190 GeV/$c^2$, in steps of 2.5 GeV/$c^2$, using no other prior knowledge of the mass. Figure \ref{topmass} shows the 2-dimensional probability density as a function of $F_0$ and $m_t$ for the data, after applying the response correction from Fig. \ref{line}. Figure \ref{results}b shows the probability density from Fig. \ref{topmass}, after integration over $m_t$. The probability in Fig. \ref{results}b is fitted to a 5$^{th}$-order polynomial as a function of $F_0$. We use the most probable output value (at the maximum) to define the extracted $F_0$. The uncertainty in $F_0$ (shaded region in Fig. \ref{results}b) is defined by the most narrow interval within which the integral of the normalized probability function contains 68.27$\%$ of the area, and reflects the statistical error convoluted with the uncertainty on the mass of the top quark: 
\begin{equation}
F_0=0.56\pm0.31({\rm stat\&m_t})
\end{equation}
This is the only uncertainty we are able to treat in this manner. The other systematic uncertainties are quite small, and were calculated by varying their impact in the Monte Carlo or data, and added in quadrature (see Table \ref{sys}). The final result is 
\begin{equation}
F_0=0.56\pm0.31({\rm stat\&m_t})\pm0.07({\rm sys}) .
\end{equation}
After combining the two errors in quadrature, the final result is $F_0$=0.56$\pm$0.31, which is consistent with expectations of the SM, as well as with the result obtained by the CDF Collaboration of 0.91$\pm$0.39 \cite{cdfwhel}. Figure \ref{result} shows our result in terms of the range of allowed angular distributions in the decay of the top quark, where $\hat\phi$ refers to the decay angle of the $l^+$ (or $d$ or $s$ quark) relative to the partner $b$ quark in the $W$ rest frame. The grey region corresponds to all possible functions with 0$\le F_0 \le$1. The 68.27\% probability interval on our measured $F_0$ restricts the allowed region to the black area, and the white central curve represents the expectation from the SM.  

\begin{table}[hb]
\caption{Impact of systematic and statistical uncertainties on the measurement of $F_0$.}
\begin{center}
\renewcommand{\arraystretch}{1.4}
\setlength\tabcolsep{5pt}
\begin{tabular}{ll}
\hline
\noalign{\smallskip}
Acceptance and linearity response & 0.055\\
Jet energy scale & 0.014\\
Spin correlations in $t\bar t$ events & 0.008\\
Parton distribution functions & 0.008\\
Model for $t\bar t$ production & 0.020\\
Multiple interactions & 0.006\\
Multijet background & 0.024\\
\hline
Total systematic uncertainties, except for $m_t$ & 0.070 \\
Statistics and uncertainty in $m_t$ & 0.306\\
\hline
Total uncertainty & 0.314 \\ 
\end{tabular}
\end{center}
\label{sys}
\end{table}

In summary, we have extracted a longitudinal-helicity fraction of 0.56$\pm$0.31 for $W$ boson decays in two lepton+jets channels in $t \bar t$ events. Although our measurement is limited by the small event sample of Run I, this powerful technique should provide far greater sensitivity to any departures from the SM in the far larger data sample anticipated in Run II.

\input{acknowledgement_paragraph_r1.tex}
\end{document}

%% file: list_of_authors_r1.tex
%
\author{                                                                      
V.M.~Abazov,$^{21}$                                                           
B.~Abbott,$^{54}$                                                             
A.~Abdesselam,$^{11}$                                                         
M.~Abolins,$^{47}$                                                            
V.~Abramov,$^{24}$                                                            
B.S.~Acharya,$^{17}$                                                          
D.L.~Adams,$^{52}$                                                            
M.~Adams,$^{34}$                                                              
S.N.~Ahmed,$^{20}$                                                            
G.D.~Alexeev,$^{21}$
% ...
\\                                                                            
\vskip 0.30cm                                                                 
\centerline{(D\O\ Collaboration)}                                             
\vskip 0.30cm                                                                 
}                                                                             
\address{                                                                     
\centerline{$^{1}$Universidad de Buenos Aires, Buenos Aires, Argentina}       
\centerline{$^{2}$LAFEX, Centro Brasileiro de Pesquisas F{\'\i}sicas,         
                  Rio de Janeiro, Brazil}                                     
\centerline{$^{3}$Universidade do Estado do Rio de Janeiro,                   
                  Rio de Janeiro, Brazil}                                     
\centerline{$^{4}$Institute of High Energy Physics, Beijing,                  
                  People's Republic of China}                                 
\centerline{$^{5}$Universidad de los Andes, Bogot\'{a}, Colombia}             
\centerline{$^{6}$Charles University, Center for Particle Physics,            
                  Prague, Czech Republic}                                     
\centerline{$^{7}$Institute of Physics, Academy of Sciences, Center           
                  for Particle Physics, Prague, Czech Republic}               
\centerline{$^{8}$Universidad San Francisco de Quito, Quito, Ecuador}         
\centerline{$^{9}$Laboratoire de Physique Subatomique et de Cosmologie,       
                  IN2P3-CNRS, Universite de Grenoble 1, Grenoble, France}     
\centerline{$^{10}$CPPM, IN2P3-CNRS, Universit\'e de la M\'editerran\'ee,     
                  Marseille, France}                                          
\centerline{$^{11}$Laboratoire de l'Acc\'el\'erateur Lin\'eaire,              
                  IN2P3-CNRS, Orsay, France}                                  
\centerline{$^{12}$LPNHE, Universit\'es Paris VI and VII, IN2P3-CNRS,         
                  Paris, France}                                              
\centerline{$^{13}$DAPNIA/Service de Physique des Particules, CEA, Saclay,    
                  France}                                                     
\centerline{$^{14}$Universit{\"a}t Freiburg, Physikalisches Institut,         
                  Freiburg, Germany}                                          
\centerline{$^{15}$Panjab University, Chandigarh, India}                      
\centerline{$^{16}$Delhi University, Delhi, India}                            
\centerline{$^{17}$Tata Institute of Fundamental Research, Mumbai, India}     
\centerline{$^{18}$CINVESTAV, Mexico City, Mexico}                            
\centerline{$^{19}$FOM-Institute NIKHEF and University of                     
                  Amsterdam/NIKHEF, Amsterdam, The Netherlands}               
\centerline{$^{20}$University of Nijmegen/NIKHEF, Nijmegen, The               
                  Netherlands}                                                
\centerline{$^{21}$Joint Institute for Nuclear Research, Dubna, Russia}       
\centerline{$^{22}$Institute for Theoretical and Experimental Physics,        
                   Moscow, Russia}                                            
\centerline{$^{23}$Moscow State University, Moscow, Russia}                   
\centerline{$^{24}$Institute for High Energy Physics, Protvino, Russia}       
\centerline{$^{25}$Lancaster University, Lancaster, United Kingdom}           
\centerline{$^{26}$Imperial College, London, United Kingdom}                  
\centerline{$^{27}$University of Arizona, Tucson, Arizona 85721}              
\centerline{$^{28}$Lawrence Berkeley National Laboratory and University of    
                  California, Berkeley, California 94720}                     
\centerline{$^{29}$California State University, Fresno, California 93740}     
\centerline{$^{30}$University of California, Irvine, California 92697}        
\centerline{$^{31}$University of California, Riverside, California 92521}     
\centerline{$^{32}$Florida State University, Tallahassee, Florida 32306}      
\centerline{$^{33}$Fermi National Accelerator Laboratory, Batavia,            
                   Illinois 60510}                                            
\centerline{$^{34}$University of Illinois at Chicago, Chicago,                
                   Illinois 60607}                                            
\centerline{$^{35}$Northern Illinois University, DeKalb, Illinois 60115}      
\centerline{$^{36}$Northwestern University, Evanston, Illinois 60208}         
\centerline{$^{37}$Indiana University, Bloomington, Indiana 47405}            
\centerline{$^{38}$University of Notre Dame, Notre Dame, Indiana 46556}       
\centerline{$^{39}$Iowa State University, Ames, Iowa 50011}                   
\centerline{$^{40}$University of Kansas, Lawrence, Kansas 66045}              
\centerline{$^{41}$Kansas State University, Manhattan, Kansas 66506}          
\centerline{$^{42}$Louisiana Tech University, Ruston, Louisiana 71272}        
\centerline{$^{43}$University of Maryland, College Park, Maryland 20742}      
\centerline{$^{44}$Boston University, Boston, Massachusetts 02215}            
\centerline{$^{45}$Northeastern University, Boston, Massachusetts 02115}      
\centerline{$^{46}$University of Michigan, Ann Arbor, Michigan 48109}         
\centerline{$^{47}$Michigan State University, East Lansing, Michigan 48824}   
\centerline{$^{48}$University of Nebraska, Lincoln, Nebraska 68588}           
\centerline{$^{49}$Columbia University, New York, New York 10027}             
\centerline{$^{50}$University of Rochester, Rochester, New York 14627}        
\centerline{$^{51}$State University of New York, Stony Brook,                 
                   New York 11794}                                            
\centerline{$^{52}$Brookhaven National Laboratory, Upton, New York 11973}     
\centerline{$^{53}$Langston University, Langston, Oklahoma 73050}             
\centerline{$^{54}$University of Oklahoma, Norman, Oklahoma 73019}            
\centerline{$^{55}$Brown University, Providence, Rhode Island 02912}          
\centerline{$^{56}$University of Texas, Arlington, Texas 76019}               
\centerline{$^{57}$Rice University, Houston, Texas 77005}                     
\centerline{$^{58}$University of Virginia, Charlottesville, Virginia 22901}   
\centerline{$^{59}$University of Washington, Seattle, Washington 98195}       
}                                                                             

%% file: acknowledgement_paragraph_r1.tex
%
We thank the staffs at Fermilab and collaborating institutions, 
and acknowledge support from the 
Department of Energy and National Science Foundation (USA),  
Commissariat  \` a l'Energie Atomique and 
CNRS/Institut National de Physique Nucl\'eaire et 
de Physique des Particules (France), 
Ministry of Education and Science, Agency for Atomic 
   Energy and RF President Grants Program (Russia),
CAPES, CNPq, FAPERJ, FAPESP and FUNDUNESP (Brazil),
Departments of Atomic Energy and Science and Technology (India),
Colciencias (Colombia),
CONACyT (Mexico),
Ministry of Education and KOSEF (Korea),
CONICET and UBACyT (Argentina),
The Foundation for Fundamental Research on Matter (The Netherlands),
PPARC (United Kingdom),
Ministry of Education (Czech Republic),
A.P.~Sloan Foundation,
and the Research Corporation.